\documentclass[prl,aps]{revtex4}

\usepackage{enumerate}
\usepackage{enumitem}
\usepackage{graphicx}
\usepackage{dcolumn}
\usepackage{bm}
\usepackage{graphicx}
\usepackage{amssymb}
\usepackage{epstopdf}
\usepackage{color}
\usepackage[usenames,dvipsnames,svgnames,table]{xcolor}
\usepackage{placeins}

\usepackage[english]{babel}
\newcommand{\LTO}{$\mbox{LaTiO}_3$ }
\newcommand{\STO}{$\mbox{SrTiO}_3$ }
\newcommand{\LAO}{$\mbox{LaAlO}_3$ }
\newcommand{\TiO}{$\mbox{TiO}_2$ }

\newcommand{\LSCO}{$\mathrm{La_{2-x}Sr_{x}CuO_{4}\;}$}
\newcommand{\YBCO}{$\mathrm{YBa_{2}Cu_{3}O_{7}\;}$}

\newcommand{\VG}{$V_{\mathrm{G}}$ }

\definecolor{orange}{rgb}{1,0.5,0}

\FloatBarrier

\begin{document}

\title{Density driven fluctuations in a two-dimensional superconductor}

\author{S. Hurand$^1$, J. Biscaras$^1$, N. Bergeal$^1$, C. Feuillet-Palma$^1$, G. Singh$^1$, A. Jouan$^1$, A. Rastogi$^2$, A. Dogra$^3$, P. Kumar$^3$, R. C. Budhani$^2$$^,$$^3$, N. Scopigno$^4$, S. Caprara$^4$, M. Grilli$^4$, J. Lesueur$^1$}

\affiliation{$^1$LPEM UMR8213/CNRS - ESPCI ParisTech - UPMC PSL Research University, 10 rue Vauquelin - 75005 Paris, France}
\affiliation{$^2$Condensed Matter - Low Dimensional Systems Laboratory, Department of Physics, Indian Institute of Technology Kanpur, Kanpur 208016, India}
\affiliation{$^3$National Physical Laboratory, New Delhi - 110012, India }
\affiliation{$^4$Dipartimento di Fisica Universit\`{a} di Roma
"Sapienza" and ISC-CNR, piazzale Aldo Moro 5, I-00185 Roma, Italy}

\date{\today}

\maketitle

\indent

\textbf{In the vicinity of a phase transition, the order parameter starts fluctuating before vanishing at the critical point. The fluctuation regime, i.e. the way the ordered phase disappears, is a characteristics of a transition, and determines the universality class it belongs to. This is valid for thermal transitions, but also for zero temperature Quantum Phase Transitions (QPT) where a control parameter in the Hamiltonian drives the transition.}

\textbf{In the case of superconductivity, the order parameter has an amplitude and a phase, which can both fluctuate according to well identified scenarios. The Ginzburg-Landau theory and its extensions describe the fluctuating regime of regular metallic superconductors, and the associated dynamics of the pair amplitude and the phase. When the system is two-dimensional and/or very disordered, phase fluctuations dominate.}

\textbf{In this article, we address the possibility that a new type of fluctuations occurs in superconductors with an \emph{anomalous dynamics}.
In particular we show that the superconducting to metal QPT that occurs upon changing the gate voltage in two-dimensional electron gases at \LAO/\STO and \LTO/\STO interfaces displays anomalous scaling properties, which can be explained by \emph{density driven superconducting critical fluctuations.}}

\textbf{More precisely, a Finite Size Scaling (FSS) analysis reveals that the product $z\nu$ ($\nu$ is the correlation length exponent and $z$ the dynamical critical one) is  $z\nu\sim3/2$. We argue that critical superconducting fluctuations acquire an anomalous dynamics with $z=3$, since they couple to density ones in the vicinity of a spontaneous electronic phase separation, and that $\nu=1/2$ corresponds to the mean-field value. This approach strongly departs from the conventional $z=1$ scenario in disordered 2D systems based on long-range Coulomb interactions with dominant phase fluctuations.}

\textbf{A $z\nu=3/2$ exponent has been recently reported in high Tc superconductor 
\LSCO ultra-thin films upon gating, which could be explained by an analogous scenario.}

\textbf{More generally, since superconducting fluctuations may couple to other electronic degrees of freedom such as nematic, charge-order, or density
fluctuations, we anticipate that a concomitant abundance of soft critical modes will affect and characterize a whole class of two-dimensional superconductors.}

\indent

The microscopic BCS theory of superconductivity and its phenomenological version, the Ginzburg-Landau theory in mean field approximation
describe the emergence of a quantum coherent phase out of a regular Fermi liquid metal$^{1}$, with a large electronic density, a well established Fermi surface and a nearly constant density of states at the Fermi level. Here  the superconducting critical temperature $T_{c}$ relates to the density of states at the Fermi level. 
Other superconductors such as high $T_{c}$ cuprates or pnictides strongly depart from this picture, with  non-Fermi liquid behaviors. 
In dilute superconductors such as doped \STO $^{2}$, $T_{c}$  depends on the electronic density $n$, and not only on the density of states. 

In the traditional BCS-Ginzburg-Landau scheme the destruction of the ordered phase as the temperature approaches the critical
temperature  $T_c$ is ruled by classic thermal fluctuations of both the amplitude and the phase of the order parameter. The spatial coherence length $\xi$ in the fluctuating phase scales with the distance to the critical temperature with a exponent $\nu$ ($\xi\propto(T-T_c)^{-\nu}$), while the dynamics of the fluctuations slows down with an exponent  $z=2$ ($\tau\propto\xi^{z}$), $\tau$ being the characteristic lifetime of the fluctuations. This leads for instance to the standard Aslamasov-Larkin$^{3}$ corrections to the conductivity above $T_c$. The same behavior is observed for a zero temperature Quantum Phase Transition (QPT), when a parameter $K$ of the Hamiltonian is continuously changed so that superconductivity is destroyed, and the metallic phase restored beyond a critical value $K_c$$^{4}$. In dirty two dimensional superconductors, the reduced screening of the Coulomb electron-electron repulsion may lead to pair breaking$^{5}$
 or to the direct localization of the superconducting pairs 
$^{6}$, with Coulomb interactions rendering the superfluid density weaker and the conjugated phase fluctuations correspondingly softer. Thus,
phase fluctuations dominate the physics of the system, which  can be described by a spin  XY model. In this context, Fisher argued that the QPT from a superconductor to an insulator is ruled by a dynamical exponent $z=1$, since long range Coulomb interactions are restored in these strongly localizing systems$^{6}$. More recently, it has also been 
suggested that approaching the transition, disorder itself destroys coherence and inhomogeneities develop even in structurally homogeneous thin films. An intrinsic granular behavior emerges, as evidenced by local tunneling spectroscopy$^{7,8}$, and explained by theories$^{9,12}$. 
One can wonder how universal are these above presented descriptions of phase transitions in superconductors, with dynamical exponents $z=1$ or $z=2$.

The recent discovery of a two-dimensional electron gas (2DEG) at oxide interfaces$^{13}$ opened fascinating perspectives since, in addition to strong confinement, they display \emph{d}-electrons related electronic properties$^{14}$, a sizable Rashba type Spin Orbit Coupling$^{15}$ and quantum orders such as superconductivity$^{16}$ and magnetism$^{17}$. Quantum confinement of anisotropic \emph{d}-orbitals leads to a rich and complex band structure$^{18,19}$, which can be filled by electrostatic back-gating. 
A key point is that this band structure evolves with the filling through self-consistent shaping of the potential well$^{20}$. It was shown that, in some circumstances, the system displays a negative inverse electronic compressibility $\partial \mu/\partial n<0$ ($\mu$ is the chemical potential), which leads to a spontaneous phase separation between high and low electronic density regions$^{21}$. The Rashba Spin Orbit Coupling can also foster an electronic instability and a correlated phase separation, which persists as the gate voltage is changed$^{22}$. Electronic correlations related to \emph{d}-orbitals may also lead to negative compressibility as recently reported in \LAO/\STO interfaces$^{23}$ together with phase separation$^{24,25}$.  
A phase diagram emerges with an \emph{electronic phase separation region} ended by a critical point. In this article, we show that a \emph{new kind of superconducting fluctuations} takes place in the vicinity of this density critical point, with a $z=3$ dynamics.

Superconductivity at \LAO/\STO $^{16}$ and \LTO/\STO $^{26}$ epitaxial interfaces has been recently investigated. The analysis of the magnetic critical fields evidences the strong two-dimensional character of the superconducting 2DEG which has a typical thickness $th$ smaller than 10~nm, in agreement with self-consistent calculations of the electronic states in the quantum well formed at the interface$^{20}$, and a superconducting coherence length of $\xi\sim40-70~nm$ ($th<\xi$). 
With a Fermi energy in the $100-150~meV$ range$^{20,27}$, the Fermi wavelength $\lambda_F$ 
is $\sim10~nm$, that is slightly greater than the extension of the well ($\lambda_F\geq{th}$). This 2DEG is therefore an extreme 2D superconductor, with an electronic sub-band structure due to the confinement.  As the doping is changed electrostatically with a 
gate deposited at the rear side of the substrate (back gating), the superconducting temperature $T_{c}$ is modulated, and shows a dome-like dependence with the gate voltage \VG$^{20,28}$, ending with a Quantum Critical Point. As \VG is increased, electronic sub-bands with increasing energy and spatial extension are populated. We showed that superconductivity is related to the occurrence of Highly Mobile Carriers setting at the edge of the quantum well, which correspond to high energy electronic states extending farther in the \STO substrate$^{20}$. In agreement with direct superfluid density measurements on \LAO/\STO  interfaces$^{29}$, it is found that a small fraction (of the order of 10\%) of the total electron density forms the superconducting state, thereby also supporting the scenario of an inhomogeneous superconducting state$^{30,31}$.
We recently studied the superconductor-localizing metal QPT that occurs when a perpendicular magnetic field is applied on the 2DEG$^{32}$. Through a finite size scaling analysis, we evidenced that the system is formed of superconducting puddles coupled through a metallic phase. Depending on the strength of the coupling controlled by \VG and the temperature, the critical behavior is dominated by the intra- or inter-puddles phase fluctuations of the superconducting order parameter. A striking result is that the exponent $z\nu\sim3/2$ at the lowest temperatures, which departs from usual results in the literature (see Goldman's review$^{33}$).
However, recent measurements by Bollinger et al.$^{34}$ on a ultra thin high $T_{c}$ superconductor film (1 to 2 unit cells thick), also found $z\nu=3/2$ upon gating.

In this article, we analyze the QPT of the superconducting 2DEG at \LAO/\STO and \LTO/\STO interfaces when using \VG as the control parameter of the Hamiltonian. The same $z\nu\sim3/2$ exponent is found, which we interpret as a new kind of superconducting fluctuations controlled by an electronic phase separation critical point.

\indent 

Samples are grown by pulsed laser deposition of 15 unit cells of \LTO or \LAO on \TiO-terminated \STO substrate on the (001) direction (see Biscaras et al.$^{26}$, Herranz et al.$^{35}$ and Rastogi et al.$^{50}$ for details). A metallic back gate is evaporated at the rear of the \STO substrate (500 $\mu$m thick) and connected to a voltage source (\VG). Standard four probes resistance measurements are made with low current (10 nA) and low frequency (13 Hz) lock-in voltage detection.

\begin{figure}[h]
\includegraphics[width=12cm]{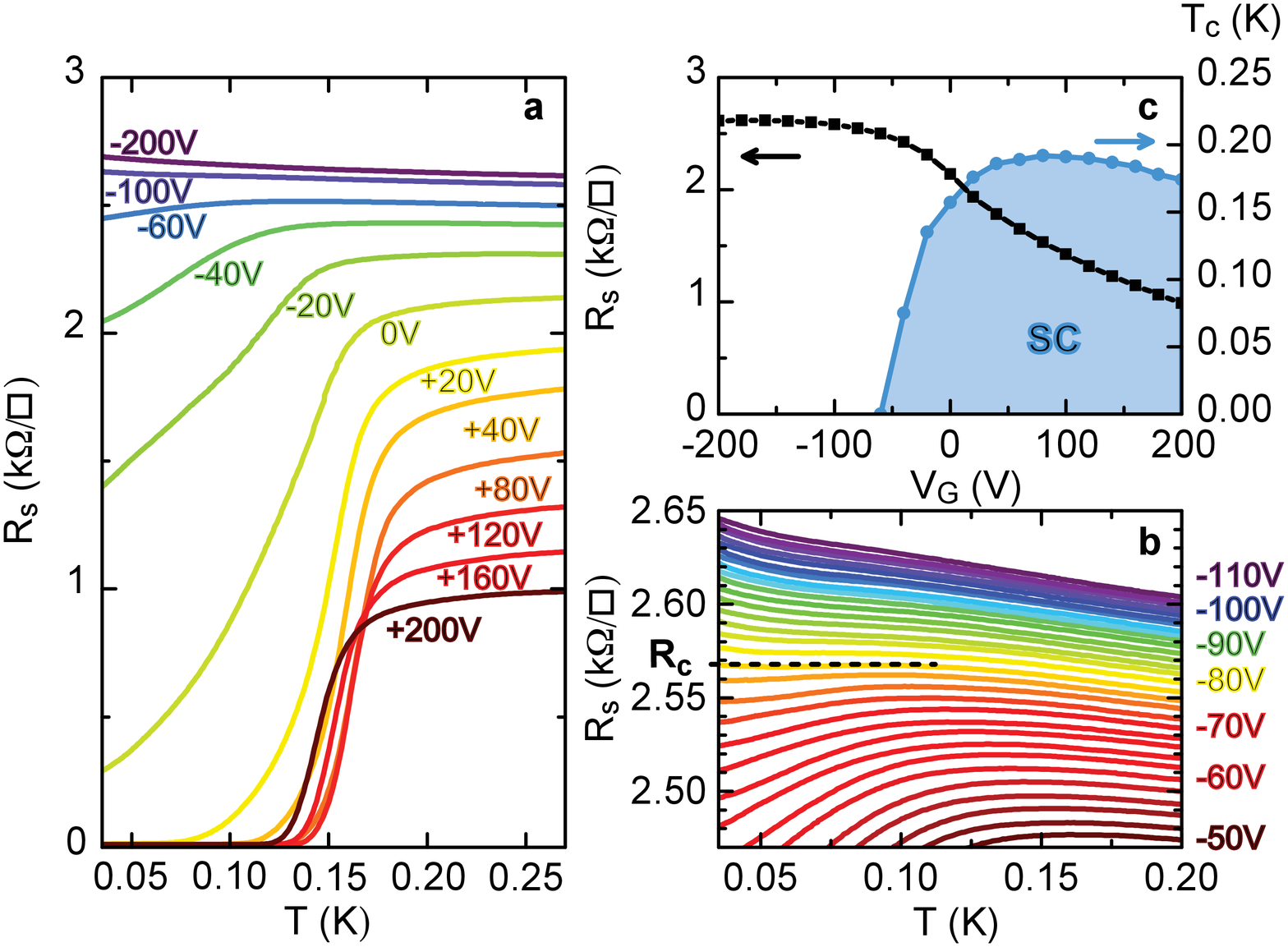}
\caption{\LAO/\STO interface : (a) Resistance per square $R_{s}$ as a function of temperature for different gate voltages \VG from $+200 V$ to $-200 V$. (b) Zoom in the low temperature region, where a plateau develops for $R_{c}\sim2.57 k\Omega/\square$. (c) Resistance per square $R_{s}$ (left axis) and superconducting critical temperature $T_{c}$ (right axis) as a function of the gate voltage \VG. The superconducting region (SC) is colored in blue.}  
\label{Figure1}
\end{figure}

\begin{figure}[h]
\includegraphics[width=12cm]{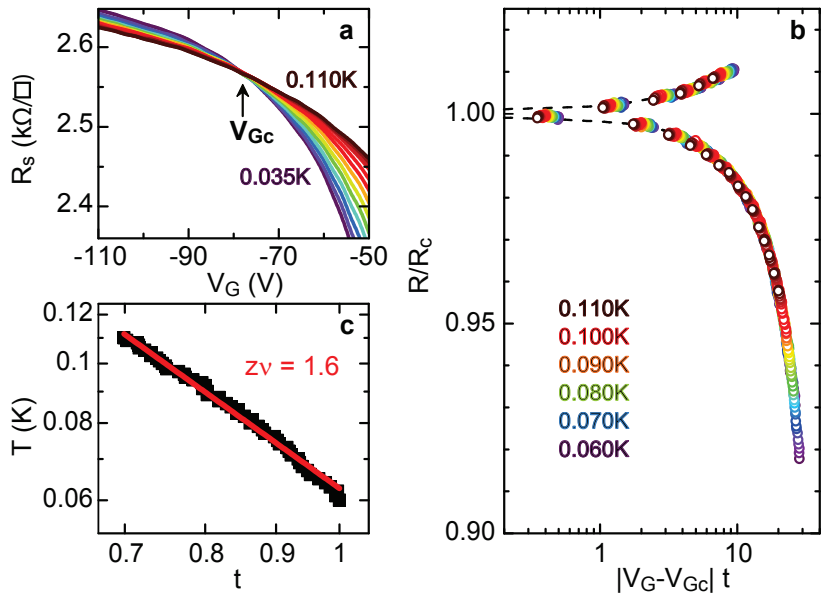}
\caption{\LAO/\STO interface : (a) $R_{s}$ as a function of the gate voltage \VG for different temperatures from 0.035 to 0.11 K. The crossing point is ($R_{c}\sim2.57 k\Omega/\square$,$V_{\mathrm{Gc}}=-78.5V$). (b) Finite size scaling plot $R_{s}/R_c$ as a function of ${\vert V_G - V_{\mathrm{Gc}} \vert t}$ (see text for the definition of $t$). (c) Temperature behavior of the scaling parameter $t$ (see text). The power law fit gives $z\nu\sim1.6\pm0.1$}
\label{Figure2}
\end{figure}

In Figure \ref{Figure1}a, the sheet resistance $R_{s}$ of a \LAO/\STO sample is plotted as a function of temperature T for different gate voltages \VG.The superconducting state is progressively destroyed in favor of a weakly localizing metal$^{26,32}$, as seen in the phase diagram Figure \ref{Figure1}c. At the lowest temperatures, a plateau develops for a critical value of the sheet resistance $R_{c}\sim2.57~k\Omega/\square$, which separates the superconducting and the normal phases. Plotting $R_{s}$ as a function of \VG for different temperatures (Figure \ref{Figure2}a) reveals a clear crossing point ($R_{c}\simeq 2.57~k\Omega/\square$, $V_{\mathrm{Gc}}=-78.5~V$) which is a sign of possible QPT. A finite size scaling analysis (FSS)$^{4}$ of the data in this region is made on Figure \ref{Figure2}b. All the sheet resistance data from 35 to 110 mK  collapse onto a single function
\begin{equation}
\frac{R_s}{R_{c}} = F\left(\frac{\vert V_{\mathrm{G}}-V_{\mathrm{Gc}} \vert}{T^{1/z\nu}}\right)
\label{eq1}
\end{equation}
if $z\nu\sim1.6\pm0.1$. This critical exponent can then be retrieved (Figure \ref{Figure2}c) by a scaling procedure$^{33}$ described in the methods section.

\begin{figure}[h]
\includegraphics[width=12cm]{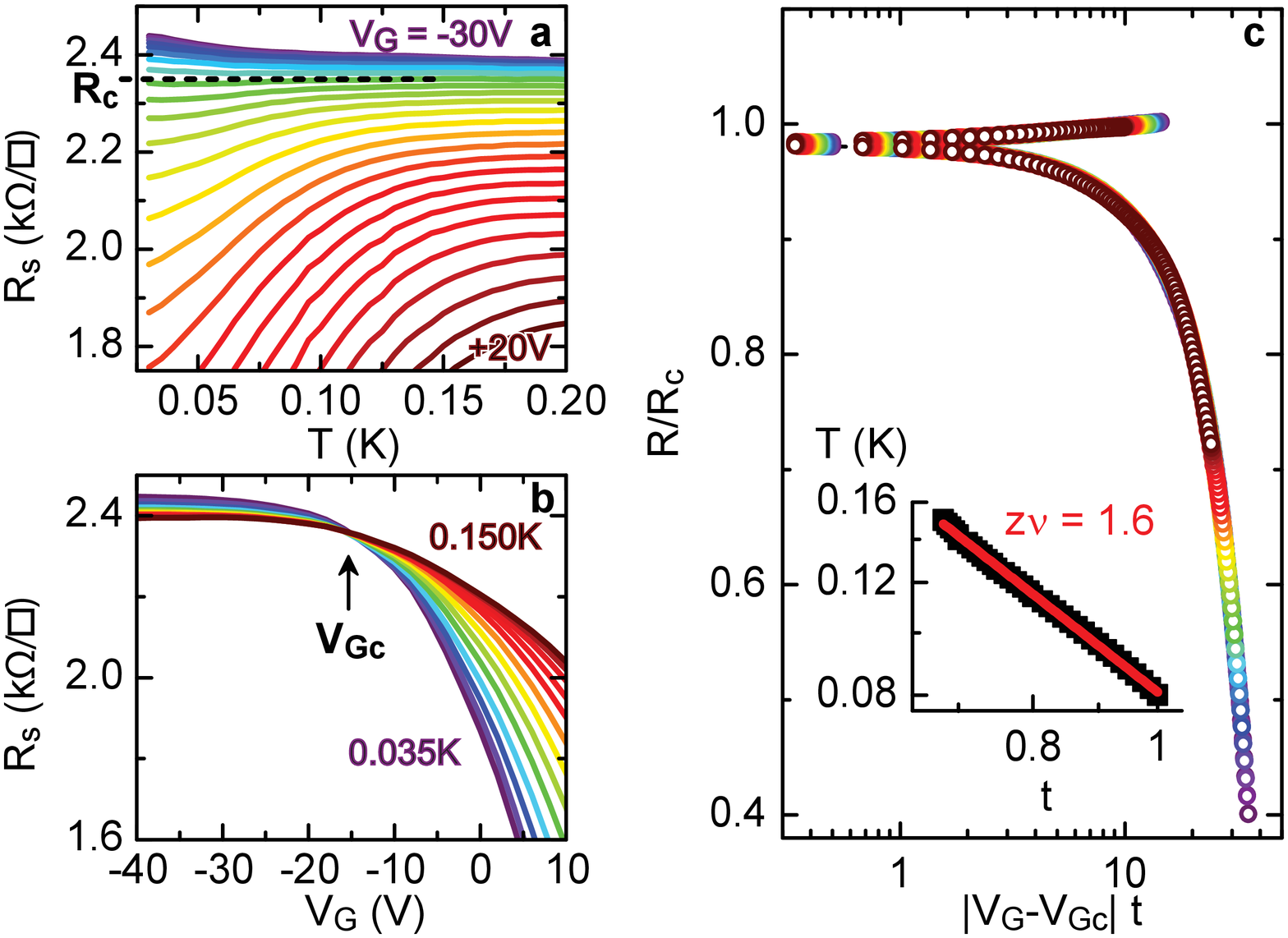}
\caption{\LTO/\STO interface : (a) Resistance per square $R_{s}$ as a function of temperature between 0.03 and 0.2 K for different gate voltages from $+30V$ to $-20V$. A plateau is observed for $R_{c}\sim2.35 k\Omega/\square$ (b) $R_{s}$ as a function of the gate voltage \VG for different temperatures from 0.035 to 0.15 K. (b)  The crossing point is ($R_{c}=\sim2.35 k\Omega/\square$,$V_{\mathrm{Gc}}=-15V$). (c) Finite size scaling plot $R_{s}/R_{c}$ as a function of  ${\vert V_G - V_{\mathrm{Gc}} \vert t}$ (see text for the definition of $t$). Inset : temperature behavior of the scaling parameter $t$ (see text). The power law fit gives $z\nu\sim1.6\pm0.1$}
\label{Figure3}
\end{figure}

The behavior of \LTO/\STO 2DEG is very similar. Figure \ref{Figure3}a displays $R_{s}$ as a function of temperature T for different gate voltages : a critical sheet resistance $R_{c}\sim2.35~k\Omega/\square$ separates the superconducting phase from the metallic one, and a crossing point is evidenced (Figure \ref{Figure3}b). The FSS analysis of the low temperature data points toward a QPT with a critical exponent $z\nu\sim1.6\pm0.1$ (Figure \ref{Figure3}c).

The exponent $z\nu\sim3/2$ is somewhat surprising, and rarely found in the literature. The most common reported values are close to $z\nu=2/3$, $z\nu=4/3$ or $z\nu=7/3$$^{33}$. Assuming a dynamical exponent $z=1$, the spatial exponent $\nu=2/3$ would therefore correspond to a clean (2+1)D XY model, coherent with a scenario where phase fluctuations dominate the behavior of the system in the vicinity of the transition, with an enhanced dimension for the quantum character of the transition. Under the same assumption $z=1$, $\nu=4/3$ and $\nu=7/3$ are typical of a classical, respectively quantum percolating behavior in highly disordered systems. The main assumption in these arguments is that the dynamical exponent $z$ equals $1$, because of enhanced long range Coulomb interactions when the system becomes insulating$^{6}$. Yazdani et al.$^{36}$ measured it in amorphous MoGe films and Markovic et al.$^{37}$ in a-Bi layers, and confirmed this value.

 Following the same assumption $z=1$, we were not able to perform a satisfying FSS with the above quoted $\nu$ exponents in none of our samples. Moreover, since the non-superconducting phase is a weakly localizing metal and not an insulator$^{26,32}$, screening effects should be substantial, and this might cast some doubts about the presence of long-range Coulomb forces customarily invoked to justify the $z=1$ critical exponent. Indeed the highly metallic character of these materials leaves the possibility open that the $z=1$ critical behavior usually invoked for the superconductor to insulator transition could be replaced by an over damped dynamics with $z>1$. Thus, in the following, we investigate an alternative scenario, in which the dynamical critical behavior of the superconducting fluctuations is imposed by the coupling to nearly critical density fluctuations with $z=3$.
In this scheme, the experimental observation of $z\nu\sim3/2$ arises from a $z=3$ dynamical critical behavior of the
superconducting fluctuations together with a mean-field like exponent $\nu=1/2$$^{38}$. 

\begin{figure}[h]
\includegraphics[width=12cm]{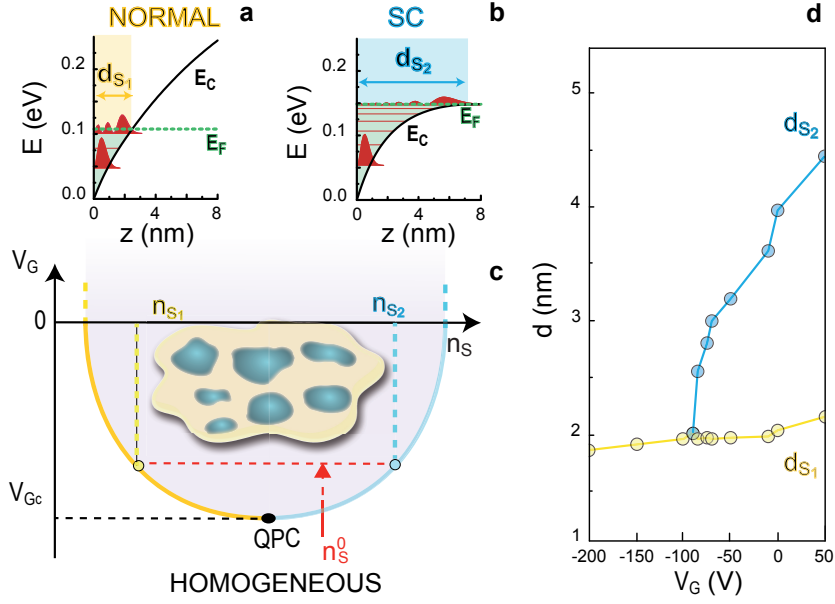}
\caption{(c) : Sketch of the electronic phase separation at oxide interfaces. A phase separation line ending by a density QCP marks out the homogeneous from the phase separated zone (in purple) in the \VG vs $n_{s}$ phase diagram. For a sample with a density $n_{s}$ (red arrow) at  \VG $< V_{\mathrm{Gc}}$, the system has an homogeneous density. When further increasing \VG, it enters the phase separated zone, and decomposes into low density $n_{s1}$ (yellow broken line) and high density $n_{s2}$ (blue broken line) droplets, as sketched within the dome. In the low density droplets, carriers are localized near the interface and remain normal (a), while in the high density ones, carriers extend far from the interface (b) and display superconductivity. The potential well presented in (a) and (b) have been calculated using Biscaras et al. method described in ref 20, for gate voltages of - 200 and + 200 V respectively. (d) : Within the Scopigno et al. theory$^{21}$, calculated gas extension $d$ as a function of \VG. Beyond $V_{\mathrm{Gc}}$ (which for the parameters chosen in the theoretical calculations occurs above $V_{\mathrm{Gc}}\approx -100 V$), $d$ remains constant for the low density regions at  $n=n_{s1}$ (yellow dots) , and increases in high density droplets at $n=n_{s2}$  (blue dots).} 
\label{Figure4}
\end{figure}

Assuming that a phase separation takes place in the system as already mentioned$^{21,22}$, we sketch in Figure \ref{Figure4}c the phase diagram of the electronic density as a function of the gate voltage. At low density (very negative \VG), the electronic system is homogeneous, and becomes phase separated when entering the instability dome. In this region, static clusters of high density $n_{2s}$ are embedded in low-density regions of density $n_{1s}$, whose proportions are given by the Maxwell construction. This intrinsic electronic inhomogeneity accounts for observed inhomogeneous superconducting properties$^{32,39}$: indeed, if $n_{2s}$ is high enough, the islands can be superconducting$^{20}$, embedded in normal zone of density $n_{1s}$. The dome ends at a Quantum Critical Point (QCP)$^{38}$, which is the analogous of the critical point of the classical liquid-gas phase diagram. In the vicinity of the QCP, critical density fluctuations are ruled by a $z=3$ dynamics. This substantially increases
the effective dimensionality $D+z$ and the system  therefore displays a mean-field exponent $\nu=1/2$$^{38}$. 
Let us first make the hypothesis that superconductivity is destroyed at the critical gate voltage $V_{Gc}$ of the QCP . In other words, this voltage corresponds to the beginning of the phase separation \emph{and} the occurrence of superconductivity (in high density droplets). In that case, superconducting fluctuations develop on density fluctuating droplets, and therefore acquire their dynamics with $z=3$$^{38}$: consequently $\nu=1/2$, and $z\nu=3/2$. 

There are good reasons for the density QCP to coincide with the occurrence of superconductivity. The first one is closely related to the physics of this system. The phase separation that takes place is likely due to non-rigid \emph{d}-wave bands and sub-bands in the quantum well. The complex band structure at the interface is still under debate, but a consensus has emerged that the shape of the well, and therefore the spacing between sub-bands together with the energy of the bottom of the conduction band change with the filling of the well, and therefore the electron density. The more electrons are added, the deeper in energy is the well. This is the origin of the non rigidity of the bands. The lowest bands are of $d_{xy}$ character, and lie close to the interface. Upon doping, the populated upper band can be of $d_{xy}$ character$^{20}$ or $d_{xz}/d_{yz}$ one$^{40}$ or even a mixture between them through Spin Orbit Coupling$^{41}$. Whatever the scenario is, the common feature is that this band strongly delocalizes within the \STO substrate when populated, which favors superconductivity$^{20,39}$. In the {framework of Ref. 21, the extension of the 2DEG as a function of gate voltage has been calculated (see Figure \ref{Figure4}d). When entering the phase separated domain, the spatial extension of the band related to $n_{s1}$ does not evolve significantly while the one related to $n_{s2}$ increases rapidly beyond 4 nm, where the superconducting phase shows up$^{20}$. Consequently, phase separation and superconductivity appear in the same region of the phase diagram. In addition, superconductivity can, by itself, favor phase separation when $T_{c}$ depends on the electronic density $n_{s}$ as it is the case here. If the pairing energy  $\Delta$ is a function of $n_{s}$, the system may gain energy by forming superconducting regions of high density in non superconducting low density ones. A Ginzburg-Landau approach extended to a diffusion model confirms this possibility, and the occurrence of an electronic instability in this case$^{38}$.

Conversely, the electronic compressibility is strongly enhanced near a density-driven superconducting critical line $T_c(n)$$^{38}$. Thus, the QCP related to the phase separation dome (where the compressibility diverges), is attracted by the superconducting QCP (where $T_c(n)$ vanishes), and the two QCP tend to merge.

Seldom have been reports of 2D superconductor to weakly localized metal or insulator transitions controlled by the electronic density so far.
Focusing on critical exponents extracted from a complete FSS analysis (see Supplementary material), a few results are found in the literature.
Parendo et al.$^{42}$ reported $z\nu\sim2/3$ in amorphous Bi films, corresponding to a clean (2+1)D XY model in this highly metallic conventional s-p compound.
In 7 u.c. thick \YBCO layers$^{43}$, $z\nu\sim2.2$ has been found corresponding to quantum percolation, which is not surprising in highly disordered thin films with up to 6 u.c. of dead layer. 
However, in 1 to 2 u.c. thick \LSCO samples, Bollinger et al. found exactly $z\nu=3/2$$^{34}$, while Garcia-Barriocanal et al. reported $z\nu$ from $1.4$ to $1.8$ for 4 u.c \LSCO thin films$^{44}$, not far from it. 
This system shares similarities with the \LAO or \LTO/\STO interfaces, such as an extreme 2D character (thickness of a few unit cells) and a low electronic density at the transition (in the $10^{13}\:e^{-}/cm^{2}$ range). It is worthwhile mentioning that the magnetic field driven transition looks also very similar in the two systems, with two critical regimes in temperature$^{32,45}$. There is no report of negative compressibility induced phase separation in \LSCO, but other intrinsic electronic inhomogeneities have been found such as fluctuating Charge Density Waves $^{46}$. The coupling of the superconducting fluctuations to other critical electronic modes near Charge Density Waves$^{47}$, stripes or fermionic nematic phases$^{48}$ QCP may lead to anomalous dynamics as well. It would be interesting to explore more widely this possibility in cuprates, Fe-pnictides or other exotic superconductors.

In conclusion, we studied the superconductor to metallic quantum phase transition in \LAO/\STO and \LTO/\STO interface as a function of the electronic density tuned by a gate voltage. The critical exponents product $z\nu\sim3/2$ is compatible with density driven fluctuations, where the superconducting fluctuations are coupled to density ones in the vicinity of an electronic phase separation critical point. This new type of fluctuations may be observed in other 2D superconductors, and perhaps also in superfluids such as $^{4}He$ on aerogels, where a similar $z\nu$ has been found$^{49}$.

\indent

The Authors gratefully thank C. Di Castro, C. Castellani, L. Benfatto and C. Kikuchi-Marrache for stimulating discussions. This work has been supported by the R\'egion Ile-de-France in the framework of CNano IdF, OXYMORE and Sesame programs, by Delegation G\'enerale \`a l'Armement (S.H. PhD Grant), by CNRS through a PICS program (S2S) and ANR JCJC (Nano-SO2DEG). Part of this work has been supported by the IFCPAR french-indian program (contract 4704-A). Research in India was funded by the CSIR and DST, Government of India. SC and MG acknowledge financial support from Sapienza University of Rome, Project AWARDS n. C26H13KZS9.

\indent
\subsection{Methods}
The resistance is rewritten as $R(\delta,t)= R_{c} F(\vert \delta \vert t)$ with $t$ an unknown parameter that depends only on $T$, and $\delta=(V_{G}-V_{Gc})$, the distance to the critical point. The parameter $t$ is then found at each temperature $T$ by optimizing the collapse around the critical point between the curve $R(\delta,t(T))$ at temperature $T$ and the curve $R(\delta,t(T_0))$ at the lowest temperature considered $T_0$, with $t(T_0)$=1. The dependence of $t$ with temperature should be a power law of the form $t= (T/T_0)^{-1/z\nu}$ in order to have a physical sense, thus giving the critical exponent product $z\nu$ . The interest of this procedure is to perform the scaling without knowing the critical exponents beforehand.

\clearpage
\subsection{Bibliography}

1.	Tinkham, M. Introduction to Superconductivity. (Dover, 2004).

2. 	Lin, X., Zhu, Z., Fauqu\'e, B.  \& Behnia, K. Fermi Surface of the Most Dilute Superconductor. \emph{Phys. Rev. X} \textbf{3}, 021002 (2013).

3.	Aslamasov, L. G.  \& Larkin, A. I. The influence of fluctuation pairing of electrons on the conductivity of normal metal. \emph{Physics Letters A} \textbf{26}, 238 (1968).

4.	Sondhi, S., Girvin, S., Carini, J. \& Shahar, D. Continuous quantum phase transitions. \emph{Rev. Mod. Phys.} \textbf{69}, 315-333 (1997).

5.	Finkel'shtein, A. M. superconducting transition temperature in amorphous films. \emph{JETP Letters (Pis'ma Zh. Eksp. Teor. Fiz. )} \textbf{45}, 46-49 (1987).

6.	Fisher, M. Quantum phase transitions in disordered two-dimensional superconductors. \emph{Physical Review Letters} \textbf{65}, 923-926 (1990).

7.	Sac\'ep\'e, B. et al. Localization of preformed Cooper pairs in disordered superconductors. \emph{Nat Phys} \textbf{7}, 239-244 (2011).

8.	Kamlapure, A. et al. Emergence of nanoscale inhomogeneity in the superconducting state of a homogeneously disordered conventional superconductor. \emph{Sci. Rep.} \textbf{3}, 2979 (2013).

9.	Bouadim, K., Loh, Y. L., Randeria, M. \& Trivedi, N. Single- and two-particle energy gaps across the disorder-driven superconductor{--}insulator transition. \emph{Nat Phys} \textbf{7}, 884-889 (2011).

10.	Seibold, G., Benfatto, L., Castellani, C. \& Lorenzana, J. Superfluid Density and Phase Relaxation in Superconductors with Strong Disorder. \emph{Physical Review Letters} \textbf{108}, 207004 (2012).

11.	Skvortsov, M. A. \& Feigelman, M. V. Superconductivity in disordered thin films: giant mesoscopic fluctuations. \emph{Physical Review Letters} \textbf{95}, 057002 (2005).

12.	Dubi, Y., Meir, Y. \& Avishai, Y. Nature of the superconductor-insulator transition in disordered superconductors. \emph{Nature} \textbf{449}, 876-880 (2007).

13.	Ohtomo, A. \& Hwang, H. Y. A high-mobility electron gas at the $LaAlO_{3}/SrTiO_{3}$ heterointerface. \emph{Nature} \textbf{427}, 423-426 (2004).

14.	Okamoto, S. \& Millis, A. Electronic reconstruction at an interface between a Mott insulator and a band insulator. \emph{Nature} \textbf{428}, 630-633 (2004).

15.	Caviglia, A., Gabay, M., Gariglio, S. \& Reyren, N. Tunable Rashba spin-orbit interaction at oxide interfaces. \emph{Physical Review} \textbf{104}, 126803 (2010).

16.	Reyren, N. et al. Superconducting interfaces between insulating oxides. \emph{Science} \textbf{317}, 1196-1199 (2007).

17.	Bert, J. A. et al. Direct imaging of the coexistence of ferromagnetism and superconductivity at the $LaAlO_{3}/SrTiO_{3}$ interface. \emph{Nat Phys} \textbf{7}, 1-5 (2011).

18.	Popovic, Z. \& Satpathy, S. Wedge-shaped potential and airy-function electron localization in oxide superlattices. \emph{Physical Review Letters} \textbf{94}, 176805 (2005).

19.	Zhong, Z., Toth, A. \& Held, K. Theory of spin-orbit coupling at $LaAlO_{3}/SrTiO_{3}$ interfaces and $SrTiO_{3}$ surfaces. \emph{Phys Rev B} \textbf{87}, (2013).

20.	Biscaras, J. et al. Two-Dimensional Superconducting Phase in $LaTiO_{3}/SrTiO_{3}$ Heterostructures Induced by High-Mobility Carrier Doping. \emph{Physical Review Letters} \textbf{108}, 247004 (2012).

21.	Scopigno, N. et al. Electronic phase separation from electron confinement at oxide interfaces. \emph{arXiv} 1506.04777v1 (2015).

22.	Caprara, S., Peronaci, F. \& Grilli, M. Intrinsic Instability of Electronic Interfaces with Strong Rashba Coupling. \emph{Physical Review Letters} \textbf{109}, 196401 (2012).

23.	Li, L. et al. Very Large Capacitance Enhancement in a Two-Dimensional Electron System. \emph{Science} \textbf{332}, 825-828 (2011).

24.	Ariando et al. Electronic phase separation at the $LaAlO_{3}/SrTiO_{3}$ interface. \emph{Nature Communications} \textbf{2}, 188-187 (2011).

25.	Pavlenko, N., Kopp, T. \& Mannhart, J. Emerging magnetism and electronic phase separation at titanate interfaces. \emph{arXiv} 1308.5319v1 (2013).

26.	Biscaras, J. et al. Two-dimensional superconductivity at a Mott insulator/band insulator interface $LaTiO_{3}/SrTiO_{3}$. \emph{Nature Communications} \textbf{1}, 89 (2010).

27.	Meevasana, W. et al. Creation and control of a two-dimensional electron liquid at the bare $SrTiO_{3}$ surface. \emph{Nat. Mat.} \textbf{10}, 114-118 (2011).

28.	Caviglia, A. et al. Electric Field Control of the $LaAlO_{3}/SrTiO_{3}$ Interface Ground State. \emph{Nature} \textbf{456}, 624 (2008).

29.	Bert, J. et al. Gate-tuned superfluid density at the superconducting $LaAlO_{3}/SrTiO_{3}$ interface. \emph{Phys Rev B} \textbf{86}, 060503 (2012).

30.	Caprara, S., Grilli, M., Benfatto, L. \& Castellani, C. Effective medium theory for superconducting layers: A systematic analysis including space correlation effects. \emph{Phys Rev B} \textbf{84}, 014514 (2011).

31.	Schneider, T., Caviglia, A. \& Gariglio, S. Electrostatically-tuned  superconductor-metal-insulator quantum transition at the $LaAlO_ {3}/SrTiO_ {3}$ interface. \emph{Phys Rev B} \textbf{79}, 184502 (2009).

32.	Biscaras, J. et al. Multiple quantum criticality in a two-dimensional superconductor. \emph{Nature Materials} \textbf{12}, 542 (2013).

33.	Goldman, A. M. Superconductor-Insulator Transitions. \emph{Int J Mod Phys B} \textbf{4}, 4081-4101 (2010).

34.	Bollinger, A. T. et al. Superconductor-insulator transition in $La_{2-x}Sr_{x}CuO_{4}$ at the pair quantum resistance. \emph{Nature} \textbf{472}, 458-460 (2011).

35.	Herranz, G. et al. High mobility in $LaAlO_ {3}/SrTiO_ {3}$ heterostructures: Origin, dimensionality, and perspectives. \emph{Physical Review Letters} \textbf{98}, 216803 (2007).

36.	Yazdani, A. \& Kapitulnik, A. Superconducting-insulating transition in two-dimensional a-MoGe thin films. \emph{Physical Review Letters} \textbf{74}, 3037-3040 (1995).

37.	Markovic, N., Christiansen, C., Mack, A., Huber, W. \& Goldman, A. Superconductor-insulator transition in two dimensions. \emph{Phys Rev B} \textbf{60}, 4320-4328 (1999).

38.	Caprara, S., Bergeal, N., Lesueur, J. \& Grilli, M. Interplay between density and superconducting quantum critical fluctuations. \emph{arXiv} 1503.05997v1 (2015).

39.	Caprara, S. et al. Multiband superconductivity and nanoscale inhomogeneity at oxide interfaces. \emph{Phys Rev B} \textbf{88}, 020504 (2013).

40.	Park, S. Y. \& Millis, A. J. Charge density distribution and optical response of the $LaAlO_ {3}/SrTiO_ {3}$ interface. \emph{Phys Rev B} \textbf{87}, 205145 (2013).

41.	Joshua, A., Pecker, S., Ruhman, J., Altman, E. \& Ilani, S. A universal critical density underlying the physics of electrons at the $LaAlO_{3}/SrTiO_{3}$ interface. \emph{Nature Communications} \textbf{3}, 1129 (2012).

42.	Parendo, K. A. et al. Electrostatic tuning of the superconductor-insulator transition in two dimensions. \emph{Physical Review Letters} \textbf{94}, 197004 (2005).

43.	Leng, X., Garcia-Barriocanal, J., Bose, S., Lee, Y. \& Goldman, A. M. Electrostatic Control of the Evolution from a Superconducting Phase to an Insulating Phase in Ultrathin $YBa_{2}Cu_{3}O_{7-x}$ Films. \emph{Physical Review Letters} \textbf{107}, 027001 (2011).

44.	Garcia-Barriocanal, J. et al. Electronically driven superconductor-insulator transition in electrostatically doped $La_{2}CuO_{4+x}$ thin films. \emph{Phys Rev B} \textbf{87}, 024509 (2013).

45.	Shi, X., Lin, P. V., Sasagawa, T., Dobrosavljevic, V. \& Popovic, D. Two-stage magnetic-field-tuned superconductor-insulator transition in underdoped $La_{2-x}Sr_{x}CuO_{4}$. \emph{Nat Phys} 10, 437-443 (2014).

46.	Ghiringhelli, G., Le Tacon, M. \& Minola, M. Long-range incommensurate charge fluctuations in $(Y, Nd)Ba_{2}Cu_{3}O_{6+ x}$. \emph{Science 337} \textbf{337}, 821 (2012).

47.	Castellani, C., Di Castro, C. \& Grilli, M. Singular quasiparticle scattering in the proximity of charge instabilities. \emph{Physical Review Letters} \textbf{75}, 4650-4653 (1995).

48.	Fradkin, E., Kivelson, S. A., Lawler, M., Eisenstein, J. P. \& Mackenzie, A. P. Nematic Fermi Fluids in Condensed Matter Physics. \emph{Annu. Rev. Condens. Matter Phys.} \textbf{1}, 153 (2010).

49.	Crowell, P. A., Van Keuls, F. W. \& Reppy, J. D. Onset of superfluidity in He4 films adsorbed on disordered substrates. \emph{Phys Rev B} \textbf{55}, 12620-12634 (1997).

50.	Rastogi, A., Kushwaha, A. K., Shiyani, T., Gangawar, A. \& Budhani, R. C. Electrically Tunable Optical Switching of a Mott Insulator-Band Insulator Interface. \emph{Adv. Mater.} \textbf{22}, 4448Ð4451 (2010).

\end{document}